\documentclass[twocolumn,showkeys,aps,prb,showpacs]{revtex4-1}
\usepackage{graphicx}
\usepackage[CJKbookmarks,dvipdfm,colorlinks,linkcolor=blue,citecolor=blue]{hyperref}

\begin{document}

\title{Tuning the  electronic structures and transport coefficients of  Janus PtSSe monolayer with biaxial strain}

\author{San-Dong Guo, Xiao-Shu Guo and Ye Deng}
\affiliation{School of Electronic Engineering, Xi'an University of Posts and Telecommunications, Xi'an 710121, China}
\begin{abstract}
Due to their great potential in electronics, optoelectronics and piezoelectronics, Janus transition metal dichalcogenide (TMD) monolayers have attracted increasing research interest, the  MoSSe of which  with sandwiched S-Mo-Se structure has been synthesized experimentally.
In this work, the biaxial strain dependence of electronic structures and  transport properties of Janus PtSSe monolayer is  systematically investigated by using generalized gradient approximation (GGA) plus spin-orbit coupling (SOC).
Calculated results show that SOC has a  detrimental effect on  power factor of PtSSe monolayer, which  can be understood by  considering SOC  effects on energy bands near the Fermi level. With $a/a_0$ from 0.94 to 1.06, the energy band gap firstly increases, and then decreases, which is due to the position change of  conduction band minimum (CBM). It is found that compressive  strain can increase the strength of conduction bands convergence by changing relative position of conduction band extrema (CBE), which can enhance n-type $ZT_e$ values. Calculated results show that compressive  strain can also induce the flat valence bands around the $\Gamma$ point near the Fermi level, which can lead to high Seebeck coefficient due to large effective masses, giving rise to better p-type  $ZT_e$ values. The calculated elastic constants with $a/a_0$ from 0.94 to 1.06  all satisfy the mechanical stability criteria, which proves that the  PtSSe monolayer is mechanically stable in the considered strain range.
Our works further enrich studies  of Janus TMD monolayers, and can  motivate farther experimental works.

\end{abstract}

\pacs{72.15.Jf, 71.20.-b, 71.70.Ej, 79.10.-n ~~~~~~~~~~~~~~~~~~~~~~~~~~~~~~~~~~~Email:sandongyuwang@163.com}

\maketitle

\section{Introduction}
After the successful exfoliation of graphene\cite{q6}, two-dimensional (2D) materials have
attracted increasing attention, and lots of 2D materials have been synthesized experimentally or predicted theoretically,
 such as TMD, group-VA, group IV-VI and group-IV  monolayers\cite{q7,q8,q9,q10,q11}.
 It is firstly proposed by Hicks and Dresselhaus in 1993\cite{q2,q3} that the low-dimensional
systems or nanostructures could be potential thermoelectric materials, and the efficiency of thermoelectric conversion can be described by the dimensionless  figure of merit\cite{s1}, $ZT=S^2\sigma T/(\kappa_e+\kappa_L)$,  where  S, $\sigma$, T, $\kappa_e$ and $\kappa_L$ are the Seebeck coefficient, electrical conductivity, working temperature,  electronic and lattice thermal conductivities, respectively.
Many studies of  heat transport properties of  2D materials have been reported such as  TMD,  orthorhombic group IV-VI and group-VA  monolayers  \cite{q12,q13,q14,q15,q16,q17}. In semiconducting TMD  monolayers, it is proved that the SOC has important effects on their electronic transport properties\cite{q20}. Strain effects on  heat transport properties of 2D materials have been widely investigated.
It is found that  tensile strain can improve thermoelectric properties of $\mathrm{ZrS_2}$ and $\mathrm{PtSe_2}$  by enhancing $S^2\sigma$ and reducing $\kappa_L$\cite{q22,q23}. The strain can effectively tune $\kappa_L$ for various kinds of 2D materials, and  the $\kappa_L$ shows monotonous increase/decrease  and  up-and-down behaviors\cite{l9,l10,l11,l12} with the increasing tensile strain.

\begin{figure}
  \includegraphics[width=7.0cm]{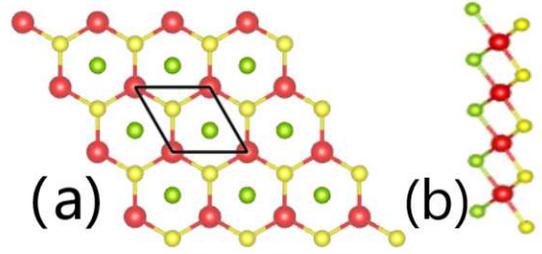}
  \caption{(Color online) The top view (a) and side view (b) crystal structure of Janus PtSSe monolayer. The Red balls represent Pt atoms, and the  yellow/green balls for S/Se atoms.}\label{t0}
\end{figure}
\begin{figure}
  \includegraphics[width=8cm]{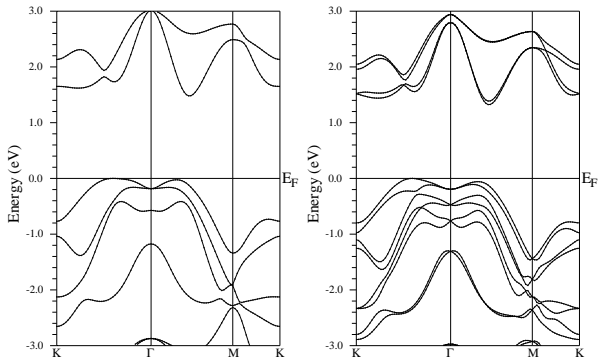}
\caption{The energy band structures  of  PtSSe monolayer  using GGA (Left) and GGA+SOC (Right). }\label{t1}
\end{figure}

Recently, Janus monolayer MoSSe has been successfully synthesised by fully replacing the top S layer with Se,
leading to an out-of-plane structural asymmetry\cite{p1}. It is found that the piezoelectric performance of monolayer and multilayer Janus TMD  MXY (M = Mo or W, X/Y = S, Se or Te) is higher than commonly used 2D materials\cite{p2}.
Although the carrier mobility of monolayer MoSSe is relatively low,  the one of bilayer or trilayer structures is quite high\cite{p2-1}.
In monolayer Janus WSeTe, the intrinsic out-of-plane built-in electric field induces a significant Rashba spin splitting\cite{p2-2}.
Janus MoSSe   monolayer may be  a potential wide solar-spectrum water-splitting photocatalyst\cite{p1-1}, and PtSSe monolayer for photocatalytic
water splitting under the visible or infrared light\cite{p1-2}.
The optical properties  of Janus MoSSe and WSSe monolayers, as well as their  vertical and lateral heterostructures,  have been investigated by the first-principles calculations\cite{p3}.
 The $\kappa_L$ of MoSSe/ZrSSe monolayer is predicted, which is much lower than that of the  $\mathrm{MoS_2}$/$\mathrm{ZrS_2}$ monolayer\cite{p4,p4-1}. The biaxial strain dependence of the electronic structures  of Janus TMD MXY (M=Mo or W, X/Y=S, Se, or Te) monolayer has been reported\cite{q5}, and the strain can effectively tune their electronic structures  and transport properties.

In this work, we systematically investigate the biaxial strain dependence of electronic structures and  transport properties of  Janus PtSSe monolayer by the first-principles calculations and Boltzmann equation. It is proved that the SOC can produce important effects on electronic structures and  transport properties of PtSSe monolayer. The energy band gap of PtSSe monolayer  shows a nonmonotonic up-and-down behavior with $a/a_0$ from 0.94 to 1.06. The compressive strain can tune the position of CBM, and induce flat valence bands, which can obviously affect S.
The  n-(p-)type S  can be enhanced  by applying  compressive strain at $a/a_0$=0.98 (0.96) point, and then the $ZT_e$ can be improved.
 In considered strain range, the PtSSe monolayer  is always mechanically stable.

\begin{table}
\centering \caption{For PtSSe monolayer, the lattice constants $a_0$ ($\mathrm{{\AA}}$), the elastic constants $C_{ij}$, shear modulus
$G^{2D}$,  Young's modulus $Y^{2D}$ in $\mathrm{Nm^{-1}}$,  Poisson's ratio $\nu$
dimensionless and the gaps with GGA and GGA+SOC. }\label{tab3}
  \begin{tabular*}{0.48\textwidth}{@{\extracolsep{\fill}}cccc}
  \hline\hline
$a_0$& $C_{11}/C_{22}$ &  $C_{12}$& $G^{2D}$\\\hline
3.66 &77.78&22.37&27.24\\\hline\hline
$Y^{2D}$& $\nu$& Gap& Gap-SOC\\\hline
71.25&0.29&1.48 &1.33\\\hline\hline
\end{tabular*}
\end{table}

\begin{figure}
   \includegraphics[width=8.0cm]{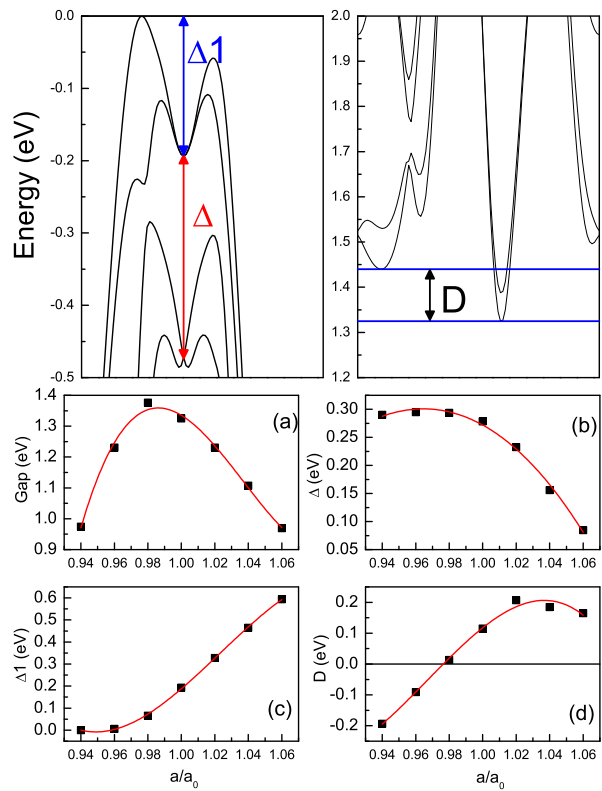}
  \caption{(Color online)The energy band structures of PtSSe monolayer near the Fermi level. The energy band gap (Gap),  $\Delta$ and $\Delta1$   at high symmetry $\Gamma$ point, and D  as a function of  $a/a_0$ by using GGA+SOC.}\label{t2}
\end{figure}

The rest of the paper is organized as follows. In the next section,
 the computational details will be given. In the third section, the strain dependence of the electronic structures and  transports  properties of  Janus PtSSe monolayer is shown. Finally, we shall give our discussions and conclusion in the fourth
section.

\begin{figure*}
  \includegraphics[width=15cm]{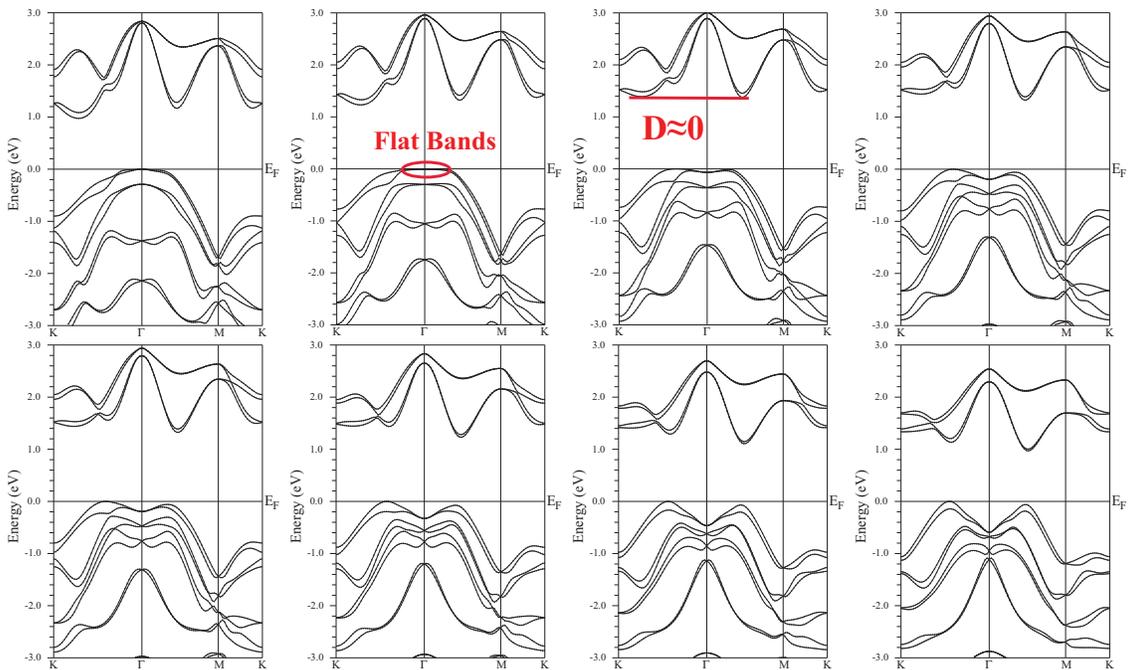}
\caption{(Color online)the energy band structures  of PtSSe monolayer with $a/a_0$ changing from 0.94 to 1.00 (Top) and 1.00 to 1.06 (Bottom) with the interval being 0.02 using GGA+SOC. }\label{t3}
\end{figure*}

\section{Computational detail}
Within the density functional theory (DFT) \cite{1}, a full-potential linearized augmented-plane-waves method
  is used to investigate electronic structures  of  PtSSe monolayer, as implemented in
the WIEN2k  package\cite{2}.
The popular GGA of Perdew, Burke and  Ernzerhof  (GGA-PBE)\cite{pbe} is used as the
exchange-correlation potential.  The  internal atomic parameters of  PtSSe monolayer are optimized with a force standard of 2 mRy/a.u..
 Due to important effects  on  electronic structure of PtSSe monolayer, the SOC was included self-consistently \cite{10,11,12,so}.
 To attain reliable results, we use  a 30 $\times$ 30 $\times$ 1 k-point meshes in the
first Brillouin zone (BZ) for the self-consistent calculation,  make harmonic expansion up to $\mathrm{l_{max} =10}$ in each of the atomic spheres, and set $\mathrm{R_{mt}*k_{max} = 8}$. The charge  convergence threshold is set as  $0.0001|e|$ per formula unit, where $e$ is
the electron charge.

By using BoltzTrap\cite{b} code, the electronic transport coefficients of PtSSe monolayer  are performed through solving Boltzmann
transport equations within the constant scattering time approximation (CSTA).
To achieve the convergence results,  the input parameter LPFAC is set to 20.
To attain  accurate transport coefficients, a 110 $\times$ 110 $\times$ 1 k-point meshes is used in the first BZ for the energy band calculation.
 For 2D material, the calculated  $\sigma$ and $\kappa_e$ depend on the length of unit cell along z direction\cite{2dl}, which  should be normalized by multiplying $Lz/d$  ($Lz$ for the length of unit cell along z direction,  and $d$ for the thickness of 2D material). However, the $d$  is not well defined like graphene.  In this work, the $Lz$=20 $\mathrm{{\AA}}$  is used as $d$.

\section{MAIN CALCULATED RESULTS AND ANALYSIS}
The structure of Janus PtSSe monolayer is shown  in \autoref{t0}, and the
three atomic sublayers can be observed  with Pt sandwiched between the S and Se layers, which  is similar to  $\mathrm{PtS_2}$/$\mathrm{PtSe_2}$ monolayer with the 1T phase.  Compared with $\mathrm{PtS_2}$/$\mathrm{PtSe_2}$ , the Janus PtSSe
monolayer lacks the reflection symmetry with respect to the central metal Pt atoms, and  the symmetry of PtSSe
(No.156) is lower than that of the $\mathrm{PtS_2}$/$\mathrm{PtSe_2}$  monolayer (No.164).
To avoid spurious interaction between neighboring layers, the unit cell  of  Janus PtSSe monolayer is built with the vacuum region of more than 18 $\mathrm{{\AA}}$. The optimized lattice constants of PtSSe (3.66 $\mathrm{{\AA}}$) with GGA  agree well with previous  theoretical values\cite{p1-2,q6-1},
which is between the ones of $\mathrm{PtS_2}$ (3.57  $\mathrm{{\AA}}$) and $\mathrm{PtSe_2}$ (3.75 $\mathrm{{\AA}}$)\cite{q20}.
Due to hexagonal symmetry,  two independent elastic
constants $C_{11}$=$C_{22}$=77.78 $Nm^{-1}$ and $C_{12}$=22.37 $Nm^{-1}$ can be calculated, and the $C_{66}$=($C_{11}$-$C_{12}$)/2=27.24 $Nm^{-1}$, which are very close to previous  theoretical values\cite{p1-2}. The 2D Young¡¯s moduli $Y^{2D}$,  shear modulus $G^{2D}$ and Poisson's ratios are calculated\cite{ela}, and are 71.25 $Nm^{-1}$, 22.37 $Nm^{-1}$ and 0.29, respectively. The related data are shown in \autoref{tab3}.

\begin{figure*}
  \includegraphics[width=15cm]{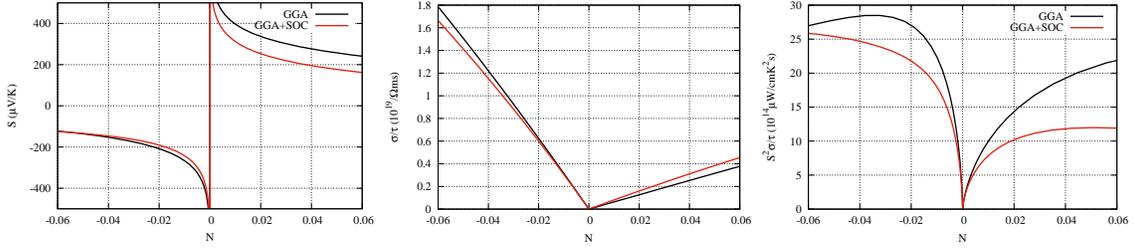}
  \caption{(Color online) the room-temperature transport coefficients (S,  $\mathrm{\sigma/\tau}$  and  $\mathrm{S^2\sigma/\tau}$) of PtSSe monolayer  as a function of doping level (N) using GGA and GGA+SOC.}\label{t4}
\end{figure*}

The SOC has  important influences on electronic structures of  TMD and  Janus TMD monolayers, and further influences their electronic transport coefficients\cite{q20,q23,q5,q21-2}. Thus, the SOC is considered for all calculations of Janus PtSSe monolayer.
The calculated energy bands for monolayer PtSSe with GGA and GGA+SOC are plotted in \autoref{t1}.
The GGA results show  the indirect gap of 1.48 eV with valence band maximum (VBM) along the $\Gamma$-K direction and CBM along the $\Gamma$-M direction. It is noted that the valence band extrema (VBE) along the $\Gamma$-M direction is very close to VBM, and the energy difference is only 0.025 eV. When the SOC is considered,  a direct gap of 1.33 eV is observed, and the energy difference between VBE along the $\Gamma$-M direction and VBM is  0.058 eV, which is larger than one with GGA. It is found that the  GGA+SOC gap is smaller than GGA one due to spin-orbital splitting, and the spin-orbit splitting at  the $\Gamma$ point ($\Delta$ in \autoref{t2}) is 0.28 eV.
Due to both
inversion and time-reversal symmetries of $\mathrm{PtS_2}$/$\mathrm{PtSe_2}$, all the bands
are doubly degenerate, but each band splits into two energy bands for PtSSe monolayer because of the lack of inversion, which is very useful  to allow spin manipulation. Our calculated GGA+SOC gaps of the $\mathrm{PtS_2}$ (1.73 eV) and $\mathrm{PtSe_2}$ (1.20 eV) monolayers\cite{q20} are very
close to the experimental values of 1.60 eV\cite{exp1} and 1.20 eV\cite{exp2}. Thus, it is reasonable to use GGA+SOC to study the electronic structures of PtSSe.

 Strain is a very effective way to tune the electronic  and phonon properties of 2D materials, and strain effects on
 energy band structures and  transport properties of TMD and Janus TMD  monolayers have been widely investigated\cite{q20,q23,q5,q21-2,q21-3,qin1}.
 Here, we examine the effects of  biaxial strain on the electronic structures and  electronic transport coefficients of   PtSSe monolayer.
The $a/a_0$ is defined to simulate biaxial strain, in which $a$ and $a_0$ are the strained and  unstrained lattice constant, respectively.
 The $a/a_0$$<$1 means  compressive strain, while  $a/a_0$$>$1 implies tensile strain.
 The energy band gap (Gap),   spin-orbit splitting value at  $\Gamma$ point ($\Delta$), the  difference  between  the Fermi level and the energy of the first valence band at  $\Gamma$ point ($\Delta1$) and the  difference  between  the second CBE   and CBM (D)  as a function of $a/a_0$   are plotted in \autoref{t2}, and the related energy band structures with $a/a_0$ from 0.94 to 1.06 are  shown  in \autoref{t3}.

 It is clearly seen that the energy band gap  firstly increases, and then decreases, when $a/a_0$ changes  from 0.94 to 1.06.  Similar phenomenon can also be found in many TMD and Janus TMD monolayers\cite{q20,q23,q5,q21-2}. With strain from compressive one to tensile one, the $\Delta$ has a slight increase, then a rapid decrease. With increasing strain, the overall trend of $\Delta$  is consistent with one of  1T TMD and Janus TMD monolayers, but is opposite to one of  2H ones\cite{q23,q5,q21-2}.
 The  strain can tune the positions of CBM and  VBM. The compressive one can change the position of CBM from one point along the $\Gamma$-M direction to another  point along the $\Gamma$-K direction, which  can be described by D. With $a/a_0$ changing from 0.94 to 1.06, the D varies from
 a negative value to a positive one, which means the change of CBM position.  The compressive one  can also tune the position of VBM from one point along the $\Gamma$-K direction to $\Gamma$ point, and can reduce the numbers of VBE from two to one. It is noted that the compressive one can
 produce the very flat valence band  around the $\Gamma$ point, for example one with $a/a_0$ being 0.96.  These can be described by $\Delta1$,and
 the $\Delta1$=0 ($\neq$0) means the one (two) VBE.
 In a word, strain can tune the position of VBM (CBM) or the numbers of VBE (CBE), and the  similar phenomenon can also be observed in  TMD and Janus TMD monolayers\cite{q20,q23,q5,q21-2,qin1}.

\begin{figure}
    \includegraphics[width=8cm]{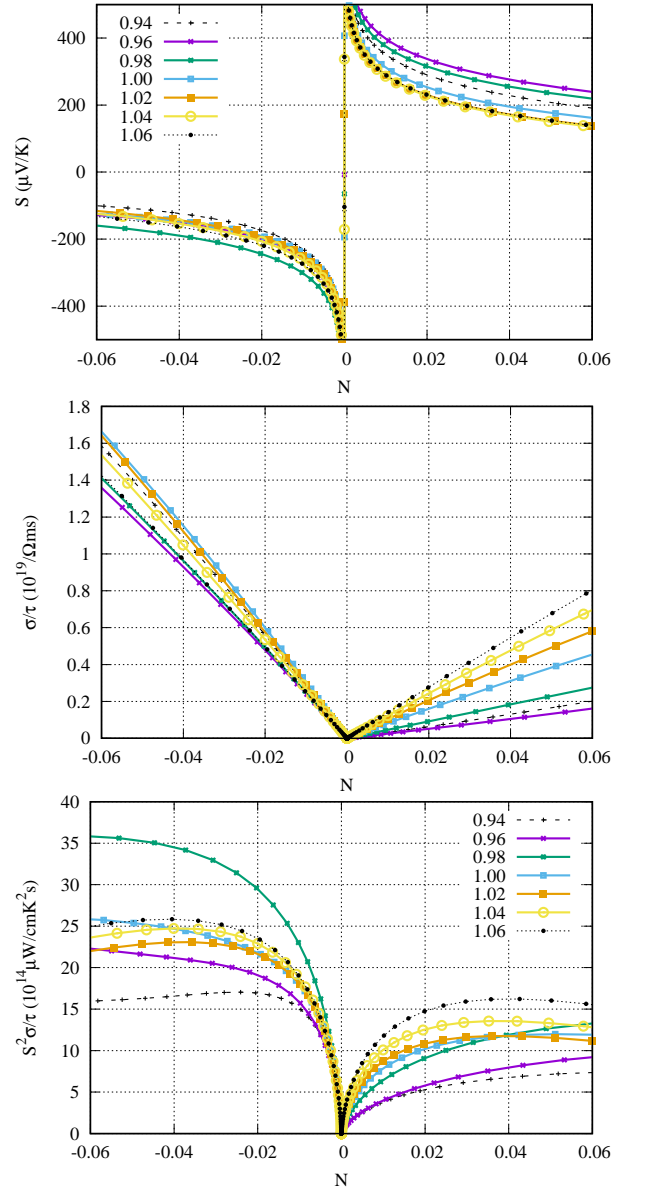}
  \caption{(Color online) the room-temperature transport coefficients (S,  $\mathrm{\sigma/\tau}$  and  $\mathrm{S^2\sigma/\tau}$) of PtSSe monolayer  as a function of doping level (N) using GGA+SOC, and  the $a/a_0$ changes  from 0.94 to 1.06  with the  interval being 0.02.  }\label{t5}
\end{figure}

 Based on CSTA Boltzmann theory within rigid band approach (RBA), the electronic  transport coefficients are investigated.
 The calculated electrical conductivity $\mathrm{\sigma/\tau}$ and  electronic thermal conductivity  $\mathrm{\kappa_e}/\tau$ depend  on scattering time $\tau$, while the  Seebeck coefficient S is independent of $\tau$. To simulate the doping effects, simply moving  the position of Fermi level within RBA is performed. The n(p)-type doping is achieved with negative (positive) doping levels by shifting  the Fermi level   into conduction  (valence) bands, giving the negative (positive) Seebeck coefficient.
 For monolayer PtSSe, the room temperature  S,   $\mathrm{\sigma/\tau}$ and  power factor with respect to scattering time $\mathrm{S^2\sigma/\tau}$  as  a function of doping level (N)   are shown in \autoref{t4},  using GGA and GGA+SOC. For 2D materials, it may be more reasonable to use electrons or holes per unit cell instead of doping concentration, which is described by N,  and the minus (positive) values mean n (p)-type doping.
  The  SOC  can induce  a detrimental influence on p-type  S of PtSSe monolayer, and  produces a slightly reduced effect on S in n-type doping.
 These can be explained  by considering SOC effects on the bands near the Fermi level. The SOC can remove the
band degeneracy   near the VBM or  CBM, leading to reduced S.
The power factor  is a comprehensive physical quantity for the
electrical performance of thermoelectric materials. Due to the
power factor being proportional to S and  $\mathrm{\sigma/\tau}$, the SOC has a remarkable detrimental influence on both n-and p-type  power factor of PtSSe in considered doping range.
It is noted that theses results also depend on the strain. When including  SOC, if the strength of bands convergence is enhanced,  the S would be improved, producing enhanced power factor. It  has been proved in  $\mathrm{WX_2}$  (X=S, Se and Te) monolayer\cite{q20}.

\begin{figure}
    \includegraphics[width=8cm]{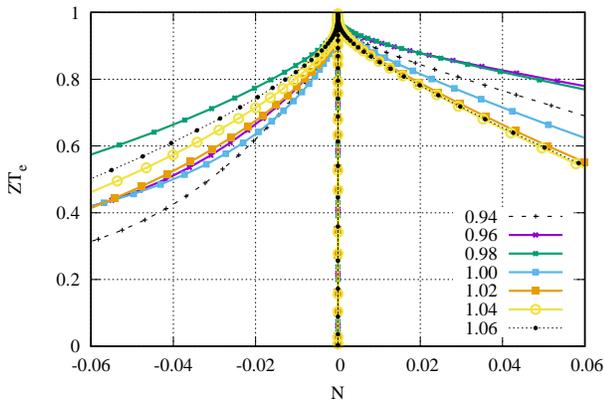}
  \caption{(Color online) the room-temperature $ZT_e$ of PtSSe monolayer  as a function of doping level (N) using GGA+SOC,  and  the $a/a_0$ changes  from 0.94 to 1.06  with the  interval being 0.02.}\label{zte}
\end{figure}

At room temperature, the biaxial strain dependence of S,  $\mathrm{\sigma/\tau}$ and $\mathrm{S^2\sigma/\tau}$ of  PtSSe monolayer    are plotted  in \autoref{t5} using GGA+SOC. Because  the electronic structures of PtSSe monolayer are  sensitively dependent on strain, the complex strain  dependence  of electronic transport coefficients are observed. With $a/a_0$ being 0.98, the largest n-type S can be attained among the considered strain points,
which can be understood  by strain-driven  accidental band degeneracies, namely bands convergence. From (d) in \autoref{t2} and \autoref{t3}, the D with $a/a_0$ being 0.98 is very close to zero, which means that the  CBE along the $\Gamma$-K direction  and CBM is almost degenerate, producing enhanced S.
Among considered  doping points, the largest p-type S  can be observed  with $a/a_0$ being 0.96, which is due to very flat valence bands around $\Gamma$ point near the Fermi level from \autoref{t3}. For metals or degenerate semiconductors, the  S is can be described by $S=\frac{8\pi^2K_B^2}{3eh^2}m^*T(\frac{\pi}{3n})^{2/3}$, in which   $m^*$, T and  $n$ is  the effective mass of
the carrier, temperature and  carrier concentration, respectively.  The flat bands can  produce very large effective mass of the carrier, and give
rise to improved S.
In  n-type doping, the largest $\mathrm{S^2\sigma/\tau}$ can be observed at $a/a_0$=0.98 point due to the largest S among the considered strain points.
For p-type doping, at $a/a_0$=0.96 point, although the largest S can be attained, the very small $\mathrm{S^2\sigma/\tau}$ can be observed, which is because the flat bands lead to very small $\mathrm{\sigma/\tau}$. Among the considered strain points, the largest p-type $\mathrm{S^2\sigma/\tau}$ can be attained at  $a/a_0$=1.06 point, which is due to the largest $\mathrm{\sigma/\tau}$.
\begin{figure}
  \includegraphics[width=8cm]{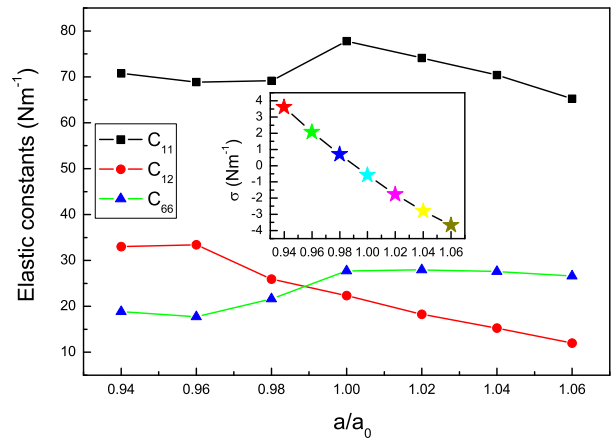}
  \caption{(Color online) The stress $\sigma$ (inset) and  elastic constants $C_{ij}$ vs  $a/a_0$ from 0.94 to 1.06 for  PtSSe monolayer.}\label{stress}
\end{figure}

The  $ZT_e=S^2\sigma T/\kappa_e$ can be defined as an upper limit of $ZT$, which  neglects the  $\kappa_L$.
At room temperature, the $ZT_e$ of Janus PtSSe monolayer  as a function of N  with $a/a_0$ changing from 0.94 to 1.06 are plotted in \autoref{zte}.
It is found that the trend of $ZT_e$ as a function of $a/a_0$ is consistent with one of S.
By the Wiedemann-Franz law: $\kappa_e=L\sigma T$ ($L$ is the Lorenz number), the   $\mathrm{\kappa_e}$ relates to   $\mathrm{\sigma}$, and then $ZT_e=S^2/L$ can be attained. Thus, the strain-improved  $\mathrm{S^2\sigma/\tau}$ caused  by enhanced S  is  beneficial to better $ZT_e$.
For example, compressive strain can significantly enhance n-(p-)type $ZT_e$ of PtSSe monolayer at $a/a_0$=0.98 (0.96) point  by compressive strain-improved  S.
It is noted that the very high $ZT$ values of many 2D materials have been reported\cite{l5-1,na}, which is because the $\kappa_e$ are calculated by the Wiedemann-Franz law with  $L$ being constant.  In our calculations, the $\kappa_e$ are used from
outputs of BoltzTrap, where the $L$ depends on temperature and doping level.

Finally,  to study the mechanical  stability of PtSSe monolayer with strain, the elastic constants $C_{ij}$ are calculated as a function of $a/a_0$,
and are plotted in \autoref{stress}. In considered strain range, they all satisfy the  Born  criteria of mechanical stability for 2D
hexagonal crystals\cite{ela}: $C_{11}>0$ and  $C_{66}>0$. The stress at different $a/a_0$ point is also calculated,  shown in \autoref{stress}. In considered strain range,  the stress is relatively small from  -3.7 $\mathrm{Nm^{-1}}$  to 3.6 $\mathrm{Nm^{-1}}$, which  can be easily realized experimentally. That means that strain-induced flat bands and strain-improved S and $ZT_e$ can be easily observed in experiment, which can stimulate further experimental works to synthesize  PtSSe  monolayer, and then investigate the strain effects on its electronic structures and transport properties.

\section{Discussions and Conclusion}
It is very important for  investigating  the electronic structures of 2D materials to chose appropriate exchange correlation potential.
Although the HSE06 can accurately calculate the gaps of many semiconductors, the GGA is more reasonable to  study the gaps of $\mathrm{PtS_2}$  and $\mathrm{PtSe_2}$. The calculated gaps ($\mathrm{PtS_2}$ (1.73 eV) and $\mathrm{PtSe_2}$ (1.20 eV))\cite{q20} with GGA are more close to the experimental values ($\mathrm{PtS_2}$ (1.60 eV) and $\mathrm{PtSe_2}$ (1.20 eV))\cite{exp1,exp2} than ones with HSE06 ($\mathrm{PtS_2}$ (2.63 eV) and $\mathrm{PtSe_2}$ (1.74 eV))\cite{q6-1}.
Thus, in this work, the GGA is used to study the electronic structures and transport coefficients of  Janus PtSSe monolayer.
For TMD and Janus TMD monolayers,  the SOC  produces  a remarkable influence on their electronic structures, and especially transport coefficients\cite{q20,q5}.
The SOC not only  can reduce the power factor of TMD and Janus TMD monolayers, but can also obviously improve  one, which depends that the strength of bands convergence  will be strengthened or weakened caused by SOC.
\begin{figure}
   \includegraphics[width=8.0cm]{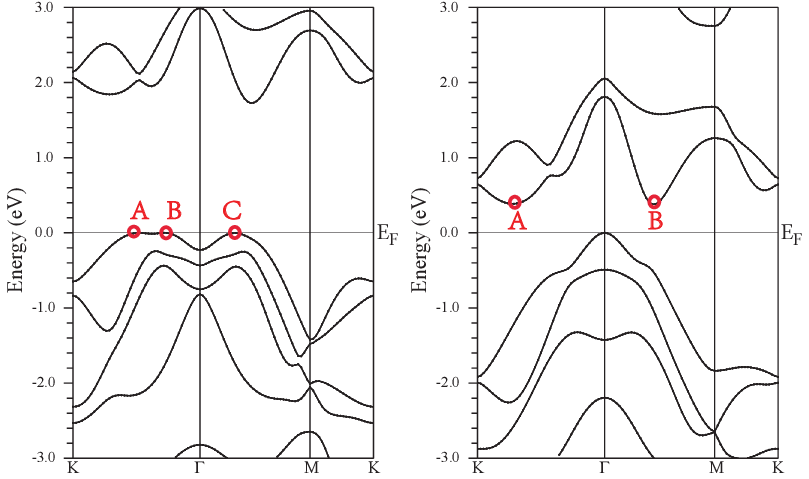}
  \caption{(Color online)The energy band structures of  $\mathrm{PtS_2}$  and $\mathrm{PtTe_2}$ using GGA+SOC.}\label{c}
\end{figure}

Strain can tune electronic structures, topological and transport properties of 2D materials by changing the distance between the atoms.
Strain can tune the positions of VBM and CBM, the numbers of VBE and CBE and the strength  of bands convergence.
For PtSSe monolayer, in n-type doing, a very small compressive  strain  ($a/a_0$=0.98) can induce the bands convergence, which gives rise to improved S. For p-type doping, a small compressive  strain  ($a/a_0$=0.96) can lead to flat bands around the $\Gamma$ point, inducing the enhanced S.
In fact, in unstrained $\mathrm{PtS_2}$  and $\mathrm{PtTe_2}$, band convergence can be realized,  and their energy band structures using GGA+SOC are plotted in \autoref{c}. For $\mathrm{PtS_2}$, the A, B and C points  in the valence bands have almost the same energy, producing the valence band convergence. For $\mathrm{PtTe_2}$, in the conduction bands, the energy  difference between the A and B points is very small, giving rise to conduction band convergence.

 In summary,  the strain dependence of  electronic structures and  transport coefficients   of Janus PtSSe monolayer are systematically studied from  the reliable first-principle calculations. It is found that the SOC can produce important effects on
energy band structures and electronic   transport coefficients of Janus PtSSe monolayer.
Calculated results show that compressive strain can change  the positions of CBM and VBM of PtSSe monolayer, and can induce flat  valence bands around the $\Gamma$ point. These changes can lead to improved S, and give rise to better $ZT_e$. In the considered strain range, the PtSSe monolayer is mechanically stable by calculated elastic constants, which satisfy  the mechanical stability criteria. Our works can stimulate further experimental works to synthesize  PtSSe  monolayer, and  will motivate farther   studies of electronic transports of other Janus monolayers.

\begin{acknowledgments}
This work is supported by the National Natural Science Foundation of China (Grant No. 11404391). We are grateful to the Advanced Analysis and Computation Center of China University of Mining and Technology (CUMT) for the award of CPU hours to accomplish this work.
\end{acknowledgments}


\begin{references}
\bibitem{q6}K. S.  Novoselov et al.,  Science \textbf{306}, 666	(2004).

\bibitem{q7}J. P. Ji,   X. F. Song, J. Z. Liu et al.,  Nat. Commun. \textbf{7}, 13352 (2016).

\bibitem{q8}S. Balendhran, S. Walia, H. Nili, S. Sriram and M.Bhaskaran, small \textbf{11},  640 (2015).

\bibitem{q9}S. L. Zhang  M. Q. Xie, F. Y. Li, Z.  Yan, Y. F. Li, E. J. Kan, W. Liu,  Z. F. Chen,  H. B. Zeng,  Angew. Chem. \textbf{128}, 1698 (2016).


\bibitem{q10}M. Chhowalla,	H. S. Shin,	G. Eda,	L. J.  Li,	K. P.  Loh	and H. Zhang, Nature Chemistry \textbf{5}, 263  (2013).


\bibitem{q11}R. X.  Fei, W. B. Li, J. Li and L. Yang, Appl. Phys. Lett. \textbf{107}, 173104 (2015).

\bibitem{q2}L. D. Hicks and M. S. Dresselhaus, Phys. Rev. B \textbf{47}, 12727  (1993).

\bibitem{q3}L. D. Hicks and M. S. Dresselhaus, Phys. Rev. B \textbf{47}, 16631(R) (1993).

\bibitem{s1} Y. Pei, X. Shi, A. LaLonde, H. Wang, L. Chen and G. J. Snyder, Nature \textbf{473}, 66 (2011).

\bibitem{q12}W. Huang, H. X. Da and G. C.  Liang,  J. Appl. Phys.  \textbf{113}, 104304 (2013).


\bibitem{q13}G. P. Li, G. Q. Ding and G. Y. Gao, J. Phys.: Condens. Matter \textbf{29}, 015001 (2017).

\bibitem{q14}G. Qin, Z. Qin, W. Fang, L. Zhang, S. Yue, Q. Yan, M. Hu and G. Su, Nanoscale \textbf{8}, 11306 (2016).

\bibitem{q15}L. M.  Sandonas,D. Teich, R. Gutierrez, T. Lorenz, A. Pecchia, G. Seifert  and G. Cuniberti, J. Phys. Chem. C  \textbf{120}, 18841 (2016).

\bibitem{q16}S. D. Guo and Y. H. Wang, J. Appl. Phys. \textbf{121}, 034302 (2017).


\bibitem{q17}D. C. Zhang, A. X. Zhang,  S. D. Guo and Y. F. Duan, RSC Adv.   \textbf{7}, 24537 (2017).

\bibitem{q20}S. D. Guo and J. L. Wang, Semicond. Sci. Tech. \textbf{31}, 095011 (2016).

\bibitem{q22}H. Y. Lv,   W. J. Lu,   D. F. Shao,  H. Y. Lub and   Y. P. Sun, J. Mater. Chem. C \textbf{4}, 4538 (2016).

\bibitem{q23}S. D. Guo, J. Mater. Chem. C  \textbf{4}, 9366 (2016).

\bibitem{l9}G. P. Li, G. Q. Ding and G. Y. Gao, J. Phys.: Condens. Matter \textbf{29}, 015001 (2017).



\bibitem{l10}H. K. Liu,  G. Z. Qin, Y. Lin and M. Hu,  Nano Lett.  \textbf{16}, 3831 (2016).

\bibitem{l11}A.  Shafique  and  Y. H. Shin, Phys. Chem. Chem. Phys. \textbf{19}, 32072  (2017).

\bibitem{l12}L. Lindsay, Wu Li, J.  Carrete, N.  Mingo, D. A. Broido  and T. L. Reinecke, Phys. Rev. B \textbf{89}, 155426 (2014).

\bibitem{p1}A. Y. Lu, H. Y. Zhu, J. Xiao et al., Nature Nanotechnology \textbf{12}, 744 (2017).

\bibitem{p2}L. Dong, J.  Lou and V.  B. Shenoy, ACS Nano \textbf{11}, 8242 (2017).


\bibitem{p2-1}W. J. Yin,  B. Wen,  G. Z. Nie and   X. L. Wei  and  L. M. Liu, J. Mater. Chem. C \textbf{6} 1693 (2018).

\bibitem{p2-2}Y. C. Cheng, Z. Y. Zhu, M. Tahir et al., Europhys. Lett. \textbf{102}, 57001 (2013).

\bibitem{p1-1}X. C. Ma,  X.  Wu,  H. D. Wang  and  Y. C. Wang, J. Mater. Chem. A  \textbf{6}, 2295 (2018).

\bibitem{p1-2}R. Peng, Y. D. Ma,  B. B. Huang and Y. Dai, J. Mater. Chem. A  \textbf{7}, 603 (2019).


\bibitem{p3}F. P.  Li, W. Wei, P. Zhao, B. B. Huang and Y.  Dai, J. Phys. Chem. Lett.   \textbf{8},  5959 (2017).

\bibitem{p4}S. D. Guo, Phys. Chem. Chem. Phys. \textbf{20}, 7236 (2018).

\bibitem{p4-1}S. D. Guo, Y. F. Li  and X. S. Guo, Comp. Mater. Sci. \textbf{161}, 16 (2019).

\bibitem{q5}S. D. Guo and J. Dong, Semicond. Sci. Tech. \textbf{33}, 085003  (2018).


\bibitem{1}P. Hohenberg and W. Kohn, Phys. Rev. \textbf{136},
B864 (1964); W. Kohn and L. J. Sham, Phys. Rev. \textbf{140},
A1133 (1965).

\bibitem{2}P. Blaha, K. Schwarz, G. K. H. Madsen, D. Kvasnicka
 and J. Luitz, WIEN2k, an Augmented Plane Wave
+ Local Orbitals Program for Calculating Crystal Properties
(Karlheinz Schwarz Technische Universit\"at Wien, Austria) 2001,
ISBN 3-9501031-1-2


\bibitem{pbe}J. P. Perdew, K. Burke and M. Ernzerhof, Phys. Rev. Lett. \textbf{77}, 3865 (1996).

\bibitem{10}A. H. MacDonald, W. E. Pickett and D. D. Koelling, J. Phys. C \textbf{13}, 2675 (1980).

\bibitem{11}D. J. Singh and L. Nordstrom, Plane Waves, Pseudopotentials and the LAPW
Method, 2nd Edition (Springer, New York, 2006).

\bibitem{12}J. Kunes, P. Novak, R. Schmid, P. Blaha and
K. Schwarz, Phys. Rev. B \textbf{64}, 153102 (2001).

\bibitem{so}D. D. Koelling, B. N. Harmon, J. Phys. C: Solid State Phys.  \textbf{10}, 3107 (1977).



\bibitem{b}G. K. H. Madsen and D. J. Singh, Comput. Phys. Commun. \textbf{175}, 67
(2006).

\bibitem{2dl}X. F. Wu, V. Varshney et al., Chem. Phys. Lett. \textbf{669}, 233 (2017).

\bibitem{q6-1}W. L. Tao, Y. Mu, C. E. Hu, Y. Cheng and G. F. Ji, Philosophical Magazine, \textbf{99}, 1025 (2019).



\bibitem{ela}R. C. Andrew, R. E. Mapasha, A. M. Ukpong and N. Chetty, Phys. Rev. B \textbf{85}, 125428 (2012).

\bibitem{q21-2}S. D. Guo, Comp. Mater. Sci. \textbf{123}, 8 (2016).

\bibitem{exp1}Y. Zhao, J. Qiao, P. Yu, Z. Hu, Z. Lin, S.P. Lau, Z. Liu, W. Ji and Y. Chai,  Adv. Mater \textbf{28}, 2399 (2016).

\bibitem{exp2}Y. Wang, L. Li, W. Yao et al., Nano Lett. \textbf{15}, 4013 (2015).


\bibitem{q21-3}E. Scalise, M. Houssa, G. Pourtois, V. Afanas'ev  and A. Stesmans,   Nano Res. \textbf{5}, 43 (2012).

\bibitem{qin1}D. Qin, X. J. Ge, G. Q. Ding, G. Y. Gao and J. T. Lv, RSC Adv. \textbf{7}, 47243 (2017).

\bibitem{l5-1}L. Cheng, H. J. Liu, X. J. Tan, J. Zhang, J. Wei, H. Y.  Lv, J.  Shi and X. F. Tang, J. Phys. Chem. C \textbf{118}, 904 (2014).

\bibitem{na} F. Q. Wang, S. Zhang, J. Yu and Q. Wang, Nanoscale \textbf{7}, 15962 (2015).


\end{references}
\end{document}